# Phenomenological analysis of the 2020 COVID-19 outbreak dynamics

Borko Stosic

Universidade Federal Rural de Pernambuco, Departamento de Estatística e Informática, Rua Dom Manoel de Medeiros s/n, Dois Irmãos, 52171-900 Recife/PE, Brazil

**Abstract**

In the wake of the COVID-19 virus outbreak, a brief phenomenological (descriptive, comparative) analysis of the dynamics of the disease spread among different countries is presented. Results indicate that the infection spread dynamics is currently the most pronounced in the USA (confirmed cases are currently doubling every 2.16 days, with a decreasing doubling time tendency), while other countries with the most confirmed cases show different values, and tendencies. The reported number of deaths is currently doubled every 2.28 days in Germany, 2.56 days in France, 2.57 days in Switzerland, 2.59 days in France, and 2.62 days in USA, while only France and USA are currently exhibiting further acceleration (diminishing doubling time).

**Introduction**

The COVID-19 virus has been spreading at an alarming rate over the world over the last few months, and it has been officially declared a pandemic by World Health Organization on March 11, 2020 [1]. Emerging in China, and initially being mostly contained in that country, as of recently the COVID-19 virus has been spreading worldwide. This dispersion has invoked stringent protective measures against its spread in most countries in the world, which have been taking place mainly over the last weeks, ranging from self-isolation [2] to curfews [3].

The current situation is precarious: different countries are implementing different measures to mitigate the propagation of the virus, but the outcome of the effectiveness of these diverse measures remains to be seen. With the aim of providing a comparative assessment of the virus spread dynamics in different countries, in this work the data provided by the Johns Hopkins University Center for Systems Science and Engineering [4] is used.

A purely phenomenological approach is adopted, without considering different measures implemented by individual countries to contain the virus spread. As more data becomes available with the progress of the pandemic, results stemming from such phenomenological studies should provide a basis for a more thorough assessment of different countries' response measures, and their comparative evaluation, in the direction of establishing the best practice guidelines.

**Data and methodological approach**

As already mentioned, the data used in this work were provided by the Johns Hopkins University Center for Systems Science and Engineering [4]. In Fig.1 the number of confirmed cases, number of deaths, and number of recovered cases, are shown as a function of time between January 22 and March 20, 2020 on semi-logarithmic scale, for the nine countries with the most confirmed cases. Linear behavior observed in different segments of these plots corresponds to an exponential growth of the corresponding variable, which may be formulated as

$$N(t) = N_0 \, 2^{t/T_d} \quad , \tag{1}$$

where $N_0$ is the number of cases at the beginning of the observed period, $t$ is the time (in days), and $T_d$ is the doubling time: the time in which the corresponding variable $N(t)$ attains double the initial value $N_0$. It can be seen from Fig. 1 that most countries go through a two stage process: after the initial spread, the number of confirmed cases saturates, and then after a given threshold (typically between 10 and a 100 confirmed cases) it starts to grow in exponential fashion (linear behavior on the plots). Linear tendency in different time segments is also observed for the other two variables, the number of deaths, and the number of recovered cases.

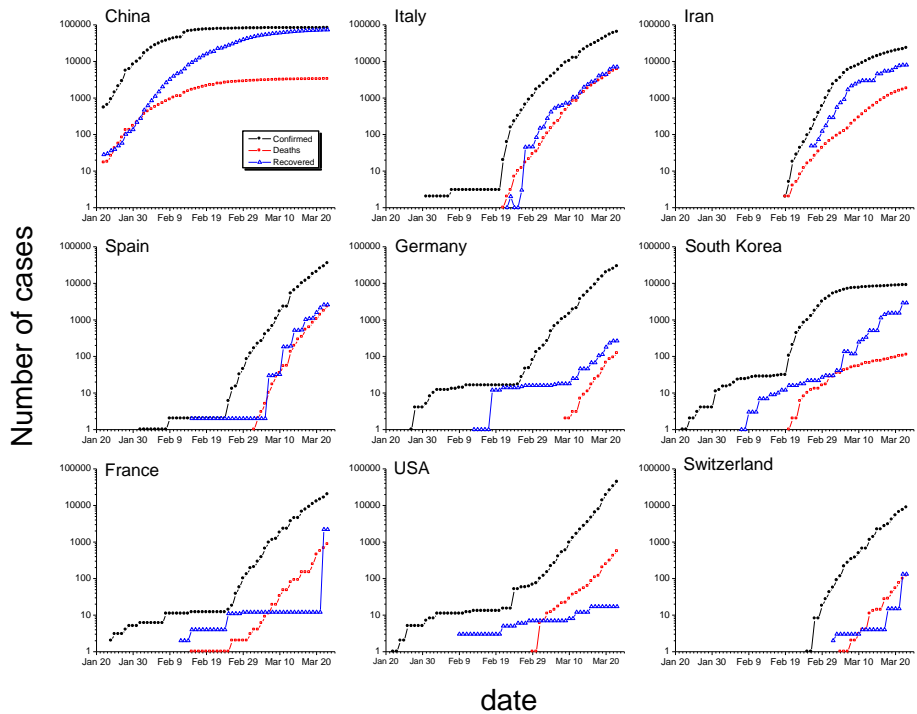

**Figure 1.** Number of confirmed cases (full circles), deaths (open squares), and recovered cases (open triangles), as a function of time, for the nine most affected countries.

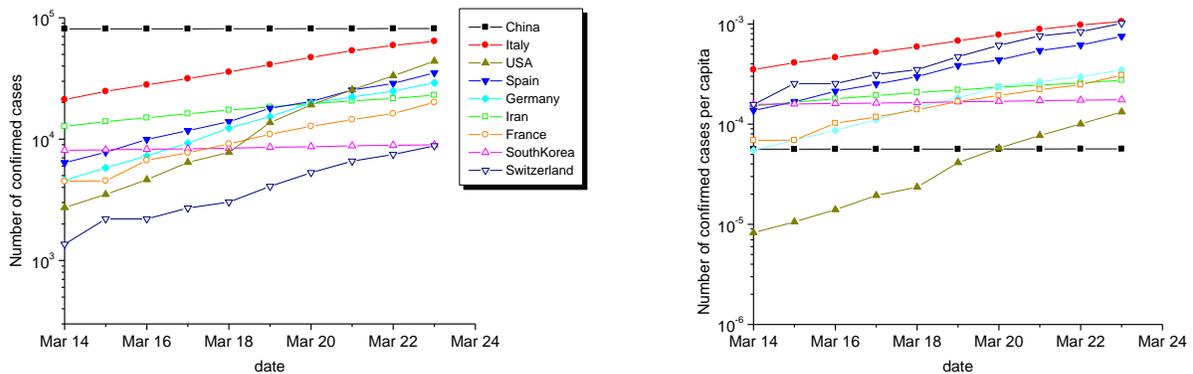

**Figure 2.** Number of confirmed cases (left) and confirmed cases per capita (right) over the last ten days, for the nine most affected countries.

The absolute number of confirmed cases, and the number of confirmed cases per capita, for the nine most affected countries, for the last ten days, are shown on Fig. 2. While the ranking of countries by the absolute number of confirmed cases is seen on the legend of Fig.2, this order is radically changed if one considers the population of these countries. As seen on the right panel of Fig. 2, the country with the largest number of confirmed cases per capita is currently Italy, followed by Switzerland, Spain, Germany, Iran, France, South Korea, USA and China.

Rather than attempt to devise a full-fledged functional model to describe the behavior of the variables in the phenomenological plots of Fig. 1, one may resort to piecewise exponential modeling: on each day, for every country, we may look at a window of size $w$ (the current day together with $w-1$ previous days), and perform least squares regression to obtain the evolution of the doubling time $T_d$. The choice of $w$ influences the results: small $w$ values result in large fluctuations of the doubling time estimate, while large values yield smoother results, but introduce latency in the estimates. To demonstrate this effect, in Fig. 3 results are shown for doubling time estimates, for number of confirmed cases in Italy, from March 1 to March 23, for different window sizes $2 \leq w \leq 10$.

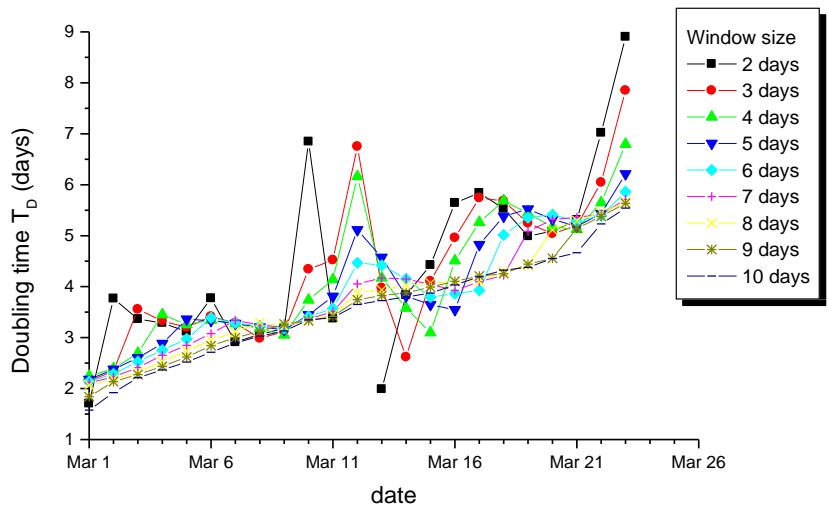

**Figure 3.** Doubling time estimate for Italy, for different choices of window size $w$.

It is seen on Fig. 3 that a choice of small $w$ values indeed leads to large fluctuations (indeed, for $w=2$ there is a missing point for March 12, as the number of confirmed cases was 12462 for both March 11 and 12, yielding infinite doubling time). As the doubling time estimates for different countries assume a rather large range of values, henceforth the choice of $w=10$ is made, although a lag of nine days is thus introduced that could obscure abrupt changes

**Results and discussion**

The confirmed cases doubling time estimates with $w=10$, for the nine most affected countries is shown in Fig. 4.

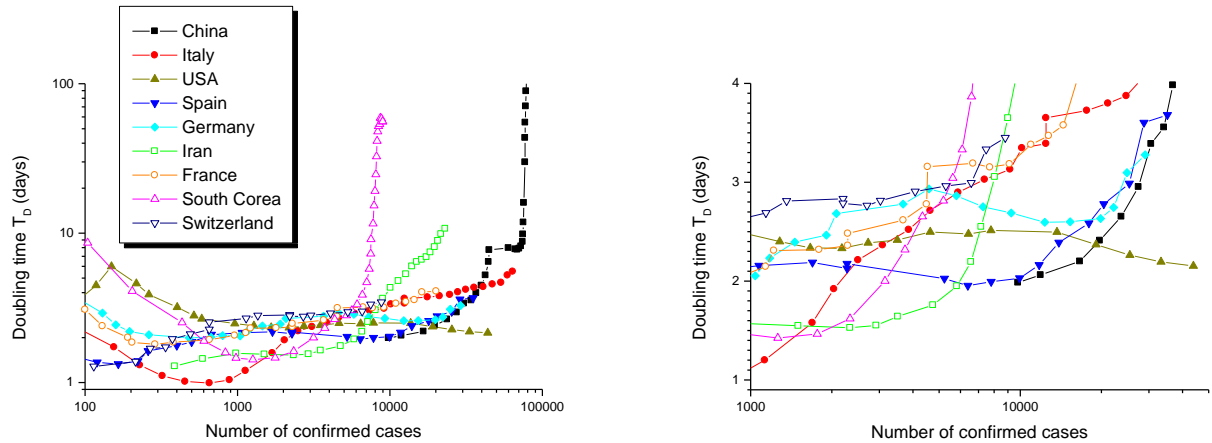

**Figure 4.** The 10-day window doubling time as a function of the number of confirmed cases, for the nine most affected countries, on double logarithmic scale. In the right panel, the same data are zoomed-in to the scale 1000-40000 confirmed cases on semi-logarithmic scale, to emphasize current tendencies.

From the results displayed in Fig. 4 it follows that Italy and USA have surpassed the growth rate of (attained lower doubling times than) China, at the similar level of confirmed cases, while all the other countries still retain higher doubling time levels. Perhaps the most attention should be drawn to the tendency observed for USA, where the doubling time has yet to assume a systematic upward trend. By far the best scenario is observed in South Korea, followed by Iran. Although Spain has previously reached doubling times lower than USA, Germany, France and Switzerland, the recent rapid doubling time increase indicates that the situation in that country is improving.

As seen on Fig.1, the death rates also follow exponential growth. The results of piecewise exponential modeling for doubling time dynamics for the nine most affected countries, are shown in Fig. 5.

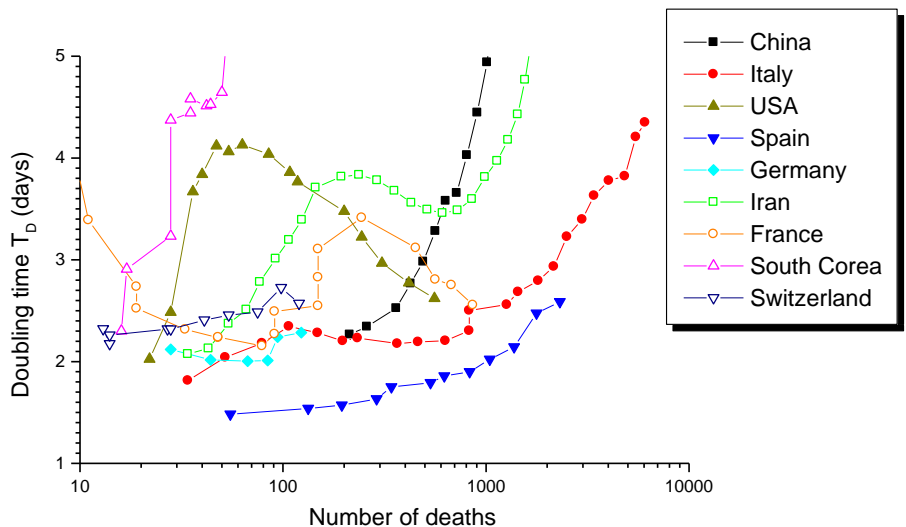

**Figure 5.** The 10-day window doubling time, as a function of the number of deaths, for the nine most affected countries.

The death doubling time appears to be worst (lowest) in Spain, followed by Italy, and Iran - all three being below the death doubling time in China at similar reported levels. Death doubling time in France and USA has been decreasing over the last days, and at current levels has also shown to be below those reported in China. Perhaps this doubling time decay shall reach a minimum, and be followed by doubling time growth, as observed in the curves of Italy and Iran, but this development remains to be seen over the next days.

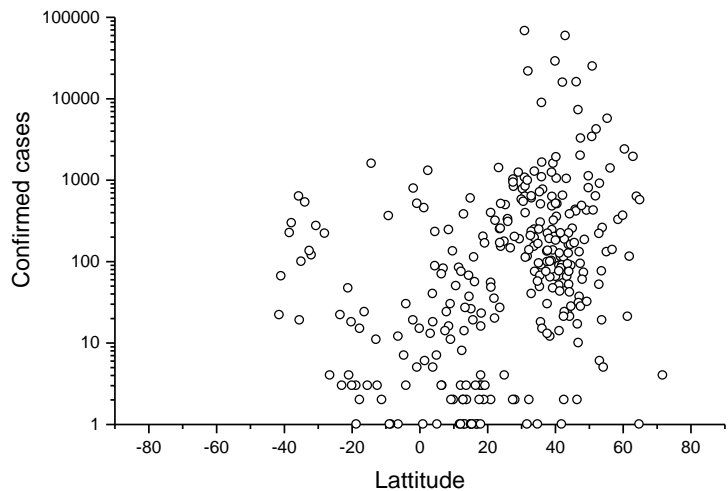

**Figure 6.** Scatter plot of confirmed COVID-19 cases on March 22, 2020, as a function of Latitude.

Finally, in Fig. 6 the number of confirmed cases on March 22, 2020 is shown on a scatterplot as a function of latitude. It is seen that the most prevalent cases are found on the north hemisphere, between 20° and 60° latitude. While one may hope that summer temperatures may alleviate the burden of the virus spread, some high level occurrences in countries south of the Equator where the summer is currently at its height, tend to diminish this hope.

**Conclusions**

The current study aims to provide a contribution to the assessment of the current observational data on the COVID-19 virus outbreak, from the purely phenomenological point of view. While the current picture is extremely volatile and may change from day to day, here an attempt is made to contribute to a better understanding of the tendencies observed in the data compiled daily by the Johns Hopkins University Center for Systems Science and Engineering [4].